\documentclass[amsmath,amssymb,twocolumn]{revtex4}
\usepackage{amssymb}
\usepackage{graphicx}
\usepackage{verbatim}
\usepackage{dcolumn}
\usepackage{bm}

\begin{document}

\title{Exact solutions of a class of $S=1$
quantum Ising spin models}

\author{ Zhi-Hua Yang$^{1}$, Li-Ping Yang$^2$, Hai-Na Wu$^{3}$,
Jianhui Dai$^{1}$, and Tao Xiang$^{2,4}$}

\address{$^1$Zhejiang Institute of Modern Physics, Zhejiang University,
Hangzhou 310027,  China\\
$^2$Institute of Theoretical Physics, Chinese Academy of
Science, P.O. Box 2735, Beijing 100080, China\\
$^3$College of Science, Northeastern University, Shengyang 110006,
China\\
$^4$Institute of Physics, Chinese Academy of Sciences, P.O. Box 603,
Beijing 100080, China }

\date{\today}

\begin{abstract}

We propose a hole decomposition scheme to exactly solve a class of
spin-1 quantum Ising models with transverse or longitudinal
single-ion anisotropy. In this scheme, the spin-1 model is mapped
onto a family of the $S=1/2$ transverse Ising models, characterized
by the total number of holes. A recursion formula is derived for the
partition function based on the reduced $S=1/2$ Ising model. This
simplifies greatly the summation over all the hole configurations.
It allows the thermodynamic quantities to be rigorously determined
in the thermodynamic limit. The ground state phase diagram is
determined for both the uniform and dimerized spin chains. The
corresponding thermodynamic properties are calculated and discussed.

\end{abstract}


\maketitle

\section{Introduction}

The phase transition driven by quantum fluctuations is one of the
fundamental issues in quantum many body systems. A number of novel
phenomena associated with the transition such as the quantum
critical behavior have been observed in a variety of condensed
matter materials \cite{Greiner,Gegenwart}. One of the prototype
model systems exhibiting the quantum phase transition is the
one-dimensional spin-1/2 Ising lattice with a transverse field,
namely the transverse Ising model
(TIM)\cite{Sondhi,Sachdev,Chakrabarti}, defined by
\begin{eqnarray}
H_{TIM}=- \sum_{j} \left( J S_{j}^zS_{j+1}^z - h S_{j}^x \right) ,
\label{hTIM}
\end{eqnarray}
where $\vec{S}_j$ is the spin operator at the site $j$ on a
one-dimensional lattice of length $L$. The transverse field $h$
introduces quantum fluctuation to the system, leading to a quantum
phase transition from the ferromagnetic/antiferromagnetic ordered to
the paramagnetic disordered states above a critical value $h_c=J/2$.
Actually the model is equivalent to a free spinless fermion system
and can be exactly solved by applying the Jordan-Wigner
transformation\cite{Lieb61,Pfeuty}. Based on the exact solution, all
physical quantities, including the ground state energy, the
low-energy excitations, the specific heat and other thermodynamic
functions can be evaluated. This provides a thorough understanding
of the quantum critical behavior of this system.

However, in real materials, the moments of atoms may be larger than
$1/2$. Despite of immense efforts in the past three decades it is
still very difficult to find exact solutions of the $S=1$ or other
higher spin quantum Ising systems. This is partly due to the
existence of the spin neutral states $S^z=0$ (called as holes
hereafter) in addition to the two spin polarized states $S^z=\pm 1$
at each site in the $S=1$ spin chain. In these models, a hole can
decay into a pair of polarized spin states, and vice verse, thus
rendering the Jordan-Winger approach invalid in exactly solving the
$S=1$ TIM.

In this paper, we study a class of one-dimensional $S=1$ quantum
Ising model, defined by the following Hamiltonian
\begin{eqnarray}
H = - \sum_{j} \left( J_j S_{j}^zS_{j+1}^z + f_j \right).
\label{eq:hQIM}
\end{eqnarray}
Where, $f_j = D_{j}^x (S_{j}^{x})^2 + D_{j}^y(S_{j}^{y})^2 +D_{j}^z
(S_{j}^{z})^2 $ is the single-ion anisotropy term with
site-dependent $D_{j}^{\alpha}$($\alpha=x,y,z$).

The above model has a classical limit where $D_{j}^x=D_{j}^y$. This
corresponds to the Blume-Capel model\cite{Blume}. So the present
model can be regarded as the quantum generalization of the
Blume-Capel model. The simplest quantum case is the uniform chain
defined by\cite{Yang}
\begin{equation}
H_{QIM}=- \sum_j \left[ J S_{j}^z S_{j+1}^z + D (S_{j}^{x})^2
\right] . \label{eq:uQIM}
\end{equation}
In two or higher dimensions, this kind of quantum Ising model with
single-ion anisotropy was studied by a number of authors, based
mainly on the mean-field approximations\cite{XFJiang,Eddeqaqi}. In
particular, the ground state of the model (\ref{eq:uQIM}) was shown
to be equivalent to the $S=1/2$ TIM defined by
Eq.~(\ref{hTIM})~\cite{Oitmaa}.  Such equivalence is also valid for
quantum Ising models with bond- and site-alternations\cite{Wu} or
geometrical frustrations, such as a fully frustrated spin-1 Ising
Delta-chain\cite{Fukumoto}.

The purpose of the present paper is to study the physical properties
of the model described by Eq. (\ref{eq:hQIM}) based on the exact
solution. The key idea in solving the proposed $S=1$ model is to
divide the total Hilbert space of the $S=1$ system into a number of
subspaces labeled by the number of holes. This is what we call the
hole decomposition scheme (HDS).  This HDS was developed in our
recent work \cite{Yang} where a recursion approach based on the HDS
is suggested for the uniform chain. In the present paper, we shall
give more comprehensive investigations for various properties of the
$S=1$ model, including the case with dimerization~\cite{note3}. In
particular, we show that for a given hole configuration, each
sub-lattice system with purely polarized spins can be exactly solved
in that case. Based on the exact solution we study how the quantum
phase transitions and thermodynamic properties are affected by the
interplay between the dimerization and the single-ion anisotropy.
Depending on the strength of dimerization, we find that the system
undergoes a quantum phase transition where the criticality is the
same as that of the uniform $S=1/2$ TIM.

We note that the spin-1/2 TIM can be realized in certain
low-dimensional magnetic materials \cite{Richter,Bitko}. For the
systems with local moments larger than $1/2$, the single-ion
anisotropy generated by crystal fields and the dimerization may also
become important\cite{Abragam}. The $S=1$ TIM with the crystal field
splitting was used to describe the ferroeletric transition in
SrTiO$_3$\cite{Yamada}. In a class of quasi-one dimensional spin
chains, such as
[Ni$_2$(Medpt)$_2$($\nu$-ox)(H$_2$O)$_2$](ClO$_4$)$_2\cdot$H$_2$O,
the magnetic Ni$^{2+}$ ion shows not only a single-ion
anisotropy{\cite{A.Escuer,J.J.B}}, but also a $S=1$ bond-alternating
pattern, where Medpt is the bis(3-aminopropyl)
methylamine\cite{Shojiro,Narumi}. Recently, cold atoms or polar
molecules in optic lattices were shown to be ideal systems to
realize various quantum spin models\cite{Micheli:07}. In particular,
the spin-1 models can be implemented by trapping polar molecules
where the spin degrees of freedom can be described by the hyperfine
vibrational states\cite{Brennen:07}.

This paper is organized as follows. In Sec.\ref{sec:model} and
Sec.\ref{sec:mapping}, we discuss some general properties of the
model and introduce the HDS. In Sec.\ref{sec:diag}, we solve exactly
the Hamiltonian in the presence of dimerization. In
Sec.\ref{sec:excitation}, we study the low energy excitation spectra
and quantum phase transitions based on the exact solutions. In
Sec.\ref{sec:thermal}, we discuss in detail the recursion method
introduced in Ref. \cite{Yang} for evaluating thermodynamic
quantities. Finally, we give a summary in Sec. \ref{sec:summary}.

\section{Hole decomposition scheme}\label{sec:model}
Let us consider a S=1 Ising lattice with single-ion anisotropy,
defined by Eq.~(\ref{eq:hQIM}). $(S_j^x, S_j^y, S_j^z)$ are the
$S=1$ spin operators at lattice site $j=1,2,...,L$, with the lattice
length $L$. The uniform classical Blume-Capel model\cite{Blume}
corresponds to the symmetric case with $D_j^x=D_j^y=D$. Because only
two of these $D_{j}^\alpha$ terms are independent, we shall mainly
consider the quantum case with $D_j^x=D_j$ and $D_j^y=D_j^z=0$,
without losing generality. The case with $D_j^z\neq 0$ will be
discussed later. We shall mainly focus on the dimerization case
where
\begin{eqnarray}
J_{2j-1}=J_1,~J_{2j}=J_2, ~D_{2j-1}=D_1,~D_{2j}=D_2.
\end{eqnarray}

By definition, one has
$
(S_j^x)^2=\frac{1}{4}(S_j^{+}S_j^{+}+S_j^{-}S_j^{-}+S_j^{+}S_j^{-}+S_j^{-}S_j^{+}).
$
So it is straightforward to show that $(S_j^x)^2$ does not couple
$S_j^z=0$ state to $S_j^z =\pm 1$ states. Thus $(S_j^x)^2$ commutes
with $(S_j^z)^2$. This leads to the following theorem:

{\it Theorem 1. When $S=1$, the total hole number operator ${\hat
N}_0=L-\sum_{j=1}^L (S_j^z)^2$ commutes with the Hamiltonian for any
site-dependent $J_j$ and $D_j$}
\begin{eqnarray}
 [{\hat N}_0, H]=0.
\end{eqnarray}
It means that the total number of holes is a conserved quantity if
$S=1$. Consequently, the eigenstates of $H$ can be classified by the
eigenvalue $p$ of ${\hat N}_0$. In the discussion below, we will
call the subsystem with $p$ holes as the $p$-th sector.

Let ${\cal H}_{p}$ be the Hilbert space of the $p$-th sector, the
total Hilbert space is then given by the sum of all the subspaces:
\begin{equation}
{\cal H}= {\cal H}_{0} \oplus {\cal H}_1\oplus \cdots \oplus {\cal
H}_{L}.
\end{equation}
A complete set of the eigenstates in the $p$-th sector forms a
sub-band of the whole spectrum. Generally the lowest eigen-energy of
the $p$-sector satisfies the following theorem:

{\it Theorem 2.  Let $E(p, L)$ be the eigen-energy, corresponding to
the eigenstate $|p,L\rangle$ of $H$, and $E_0(p, L)$ be the lowest
eigen-energy in the $p$-th sector, then the following inequality
holds}
\begin{eqnarray}
E_0(p, L)<E_0(p+1, L).
\end{eqnarray}

This relationship is an extension of the Lieb-Mattis Theorem derived
initially for the uniform systems with ferromagnetic Ising
couplings\cite{Lieb62}. It holds still no matter whether the Ising
couplings are ferromagnetic or antiferromagnetic. It indicates that
in the absence of the $D_j^z$-term the ground state always lies in
the $p=0$ sector and the energy spectrum of the system has a
hierarchy structure.

\begin{figure}[tb]
\includegraphics[width=0.9\columnwidth, bb=22 9 270 196]{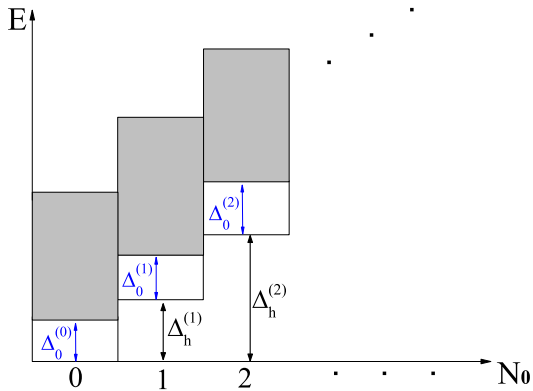}
\caption{(Color Online) The schematic picture of the energy band
hierarchy. The boxes represent the sub-bands of the corresponding
$p$-th sectors. $\Delta_0^{(p)}$ and $\Delta_h^{(p)}$ are the
minimal gaps for the fermoinic and hole excitations respectively in
the $p$-sector. Several explicit results of these gaps will be given
in Section IV.} \label{fig:band}
\end{figure}

There are two kinds of excitations in the system. One is the
fermionic excitation within a given sector. The corresponding
excitation energy is defined by $E(p,L)-E_0(p,L)$. The other is the
hole excitation and the excitation energy with respect to the ground
state is given by $E(p,L)-E_0(0,L)$. There are two kinds of minimal
excitation gaps corresponding to these excitations
\begin{eqnarray}
\Delta_0^{(p)} & \equiv & E_1(p,L)-E_0(p,L),\\
\Delta_h^{(p)} & \equiv & E_0(p, L)-E_0(0, L) \nonumber.
\end{eqnarray}
Fig.~\ref{fig:band} shows schematically the hierarchy band structure
of the system. Within each bands (or each boxes shown in
Fig.~\ref{fig:band}) there are fermionic excitations, with the
minimal gaps $\Delta_0^{(p)}$. While the minimal gaps of the hole
excitations $\Delta_h^{(p)}$ increase with $p$ when the longitudinal
anisotropy $D_j^z=0$.

\section{Mapping onto the spin-1/2 TIM's}\label{sec:mapping}

The eigenstates of the Hamiltonian Eq.~(\ref{eq:hQIM}) can be
generally expressed as $
|\Psi\rangle=\sum_{m_j}F_{m_1,m_2,\cdots,m_L} |m_1, m_2, \cdots
m_L\rangle$, with $F_{m_1,m_2,\cdots,m_L}$ being the wave function.
As the total spin and its $z$-component are not conserved, the
summation runs over all $m_j=0,\pm 1$ states. However, by Theorem 1,
the local holes are good quantum numbers and can be regarded as
non-magnetic local impurities embedded in the $S=1/2$ Ising system.
Assuming in the $p$-th sector the holes are located at $x_n$
($n=1,2,\cdots,p$), then the corresponding eigenstates can be
expressed as $
|\Psi_{(p)}\rangle=\sum_{\tilde{m}_j}F^{(p)}_{\tilde{m}_1,\tilde{m}_2,\cdots,
\tilde{m}_L} \prod_{j=1}^L\otimes|\tilde{m}_j\rangle, \label{statep}
$ where $\tilde{m}_j=\pm 1$ if $j\neq x_n$, and $\tilde{m}_{x_n}=0$.

In the $p=0$ sector, there are only two spin states at each site,
corresponding to $\tilde{m}_j=\pm 1$ respectively. They are in
one-to-one correspondence with the states of the spin-1/2 (Pauli)
operators $\sigma^z_j$: $ S_j^z|
\tilde{m}_j\rangle\Leftrightarrow\sigma_j^z|{\tilde m}_j\rangle=
{\tilde m}_j|{\tilde m}_j\rangle $. One has then the mapping
relationship: $ S_j^{\pm}S_j^{\pm}\Rightarrow 2\sigma_j^{\pm},
~~S_j^{\pm}S_j^{\mp}-1 \Rightarrow \pm\sigma_j^z $. Thus,
$(S_j^x)^2$ acts like $(1+\sigma_j^x)/2$. The original Hamiltonian,
when acting on the $p=0$ subspace, has the following reduced form
\begin{eqnarray}
H_{(0,L)}= -\sum_j J_j \sigma_{j}^z\sigma_{j+1}^z -\frac{1}{2}\sum_j
D_j ( 1 + \sigma_{j}^x ) . \label{eq:h0}
\end{eqnarray}
This is just the spin-1/2 Ising model with bond (site)-dependent
Ising couplings and transverse fields.

Now let us turn to the $p=1$ sector. If the hole is located at the
site $x_1$, the corresponding state can be written as $
|\Psi_{(1)}\rangle=\sum_{\tilde{m}_j}F^{(1)}_{\cdots,
\tilde{m}_{x_1-1},0,\tilde{m}_{x_1+1},\cdots} |\tilde{m}_1 \cdots
\tilde{m}_{x_1-1}  0_{x_1} \tilde{m}_{x_1+1} \cdots
\tilde{m}_{L}\rangle$. Now because the bonds connecting the hole are
broken, the reduced Hamiltonian, obtained by acting the original one
on the $p=1$ sector, is given by
\begin{eqnarray}
H_{(1,L)} = H'_{(0,x_1-1)}+H'_{(0,L-x_1)}-D_{x_1}. \label{h1}
\end{eqnarray}
Where $H'_{(0,l)}$ is the Hamiltonian of the spin-1/2 TIM segment of
lengths $l$ imposed by the open boundary condition. If
$|\psi^{'}(x_1-1)\rangle$ and $|\psi^{'}(L-x_1)\rangle$ are the
eigenstates of the segments $H'_{(0,x_1-1)}$ and $H'_{(0,L-x_1)}$,
respectively, then
\begin{equation}
|\Psi_{(1)}\rangle = |\psi  (x_1-1)\rangle \otimes |0_{x_1}\rangle
\otimes |\psi (L-x_1) \rangle .
\end{equation}
is the eigenstate of $H_{(1,L)}$. This HDS can be easily generalized
to the multi-hole sectors. For instance, the reduced Hamiltonian in
a $p$-hole sector is given by
\begin{eqnarray}\label{hp}
H_{(p,L)}& = & H^{'}_{(0,x_1-1)}+H^{'}_{(0,x_2-x_1-1)}+\cdots \nonumber \\
&& +H^{'}_{(0,L-x_{p})}- \sum _{n=1}^{p}D_{x_n},
\end{eqnarray}
where $x_{n}$'s are the positions of holes. Similarly, the
eigenstate of $H_{(p,L)}$ can then be expressed in terms of those of
the individual segments.

Therefore, all the eigenstates of the $S=1$ QIM (\ref{eq:hQIM}) can
be obtained by solving a set of the spin-1/2 TIMs. For the $p$-th
sector, the number of the S=1/2 TIM segments are $p+1$ or $p$,
depending on whether the original chain is periodic or open. Note
that the hole positions may vary along the chain, so there are many
different hole configurations in a given sector. The decomposition
of the total Hilbert space into the sum of subspaces can be formally
represented by $ [\textbf{2}\oplus \textbf{1}]^{\otimes L}=
\textbf{2}^{\otimes L}\oplus \textbf{2}^{\otimes (L-1)}\otimes
\textbf{1}\oplus \textbf{2}^{\otimes (L-2)}\otimes
\textbf{1}^{\otimes 2}\oplus \cdots \oplus \textbf{1}^{\otimes L}$,
where the dimension of the $S=1$ system is given by $ \mathrm{dim}
{\cal H}=(2+1)^L=\sum_{p=0}^L 2^{L-p}C_p^L=\sum_{p=0}^L \mathrm{dim}
{\cal H}_p. $

\section{dimerized spin chain}\label{sec:diag}

\subsection{The non-hole sector} \label{sub:n0}

In this section, we follow the standard approach introduced in
Refs.~\cite{Lieb61,Schultz} to diagonalize the $p=0$ sector of the
S=1 model, Eq.(\ref{eq:h0}). We introduce the fermion operators
$c_j$ and $c_j^\dagger$ and use the Jordan-Wigner transformation to
rewrite Eq.(\ref{eq:h0}) as follows
\begin{eqnarray} \label{eq:h03}
H_{(0,L)}&=& \sum_{j,q} \left[ c_j^{\dagger}A_{jq}c_q
+\frac{1}{2}(c_j^{\dagger}B_{jq}c_q^{\dagger}-c_jB_{jq}c_q)
\right]\nonumber\\
&~~& + H_{PB},
\end{eqnarray}
where $ H_{PB}=J_L(c_L^{\dagger}c_1^{\dagger}+c_L^{\dagger}c_1-
c_Lc_1^{\dagger}-c_Lc_1)(K_{L+1} +1)$, $A_{jq}=A_{qj}=
-D_j\delta_{j,q}-J_{j}\delta_{j+1,q}-J_{j-1}\delta_{j-1,q}$, and
$B_{jq}=-B_{qj}=-J_j\delta_{j+1,q}+J_{j-1}\delta_{j-1,q}$. Notice
that $H_{PB}$ is the boundary term (which disappears for the open
chains) and can be neglected in the thermodynamics limit.

Next, we introduce the Bogoliubov transformation as in the
following:
\begin{eqnarray}
\eta_k=\sum_{j} (g_{kj} c_j+h_{kj} c_j^{\dagger});
\eta_k^{\dagger}=\sum_{j} (g_{kj}^* c_j^{\dagger}+h_{kj}^* c_j).
\label{eq:eta}
\end{eqnarray}
Where, $\eta_k$ and $\eta_k^{\dagger}$ are the fermionic
quasi-particle operators with quasi-momentum $k$. They satisfy the
usual anti-commutation relations. While, $g_{kj}$ and $h_{kj}$ are
coefficient matrices, which should be complex in general. Then,
$H_{(0,L)}$ becomes
\begin{equation} \label{h-quasi}
H_{(0,L)}=\sum_{k}\Lambda(k)\eta_k^{\dagger}\eta_k
+\frac{1}{2}\sum_j A_{jj}- \frac{1}{2}\sum_k\Lambda(k).
\end{equation}

If we define $(\Phi_k)_j=g_{kj}+h_{kj}$, $(\Psi_k)_j=g_{kj}-h_{kj}$,
the eigenvalue $\Lambda(k)$ can be solved by
\begin{eqnarray}
M_{jq}(\Phi_{k})_q= \Lambda^2(k)(\Phi_k)_j; M_{jq}
(\Psi_{k})_q=\Lambda^2(k)(\Psi_k)_j,\label{eq:psi}
\end{eqnarray}
with $M$ being a symmetric matrix defined by $M=(A-B)(A+B)$, or
$M_{jq}=(D_j D_q + 4J_{j-1}J_{q-1})\delta_{j,q}+2D_q J_{j-1}\delta
_{j-1,q}+ 2D_j J_{q-1}\delta_{j,q-1}$. For the dimerized (or
alternating) chain, the $M$-matrix depends on the boundary
conditions and the parity of the chain length (even or odd). For
simplicity, we here consider the case of periodic chain with even
$L$, where
\begin{equation}\label{eq:M}
M=
\begin{pmatrix}
a_1    & b_1   & 0        &\cdots   & 0    & b_2   \\
b_1    & a_2   & b_2     &\cdots   & 0    & 0\\
0      & b_2   & a_1    &\cdots   & 0    & 0\\
\cdots &\cdots &\cdots&\cdots&\cdots& \cdots\\
0      & 0     & 0    &\cdots  & a_1  & b_1   \\
b_2    & 0     & 0    &\cdots   & b_1  & a_2
\end{pmatrix},
\end{equation}
with $a_1=D_1^2+4J_2^2$, $b_1=2D_1J_1$, $a_2=D_2^2+4J_1^2$, and
$b_2= 2D_2J_2$.

In order to solve Eq.(\ref{eq:psi}) with this $M$-matrix, we employ
the following Ansatz
\begin{eqnarray}\label{wavedimerPBC}
\begin{split}
(\Phi_k)_{2j}~~~&=A_{e}\exp[ik(2j)], \\
(\Phi_k)_{2j+1}&=A_{o}\exp[ik(2j+1)].
\end{split}
\end{eqnarray}
The ratio $\tau = A_o / A_e$ is then a measure of dimerization,
determined by
\begin{eqnarray}\label{eq:tau}
\tau=\frac{a_1 -a_2 \pm W}{2[b_1 e^ {-ik}  +b_2e^ {ik} ]},
\end{eqnarray}
where
\begin{eqnarray}\label{eq:W}
W\equiv\sqrt{(a_2-a_1)^2+4(b_1-b_2)^2+16b_1b_2\cos^2k}.
\end{eqnarray}
Note that when $a_1=a_2$ and $b_1=b_2$, we have $\tau=\pm 1$, as it
should be for the uniform chain.

The eigenvalue $\Lambda^2(k)$ is then obtained by $\Lambda^2(k)=
\Gamma^2 \pm \frac{W}{2}$, so that one has four branches of
quasi-particle excitations:
\begin{eqnarray}
\Lambda_{\pm r}(k)&=& \pm\Gamma\sqrt{1 -(-1)^r\sqrt{1
-P+Q\cos2k}}~,\label{Lamb1}
\end{eqnarray}
with $r=1,2$, $\Gamma^2  =\frac12 (D_1^2+D_2^2+4J_1^2+4J_2^2)$,
$P\Gamma^4=D_1^2D_2^2+16J_1^2J_2^2$, $Q\Gamma^4=8D_1D_2J_1J_2$, and
$k=2\pi m/L$, $(m =-L/2,\cdots L/2-1)$.

\begin{figure}[tb]
\includegraphics[width=0.9\columnwidth, bb=13 13 284 220]
{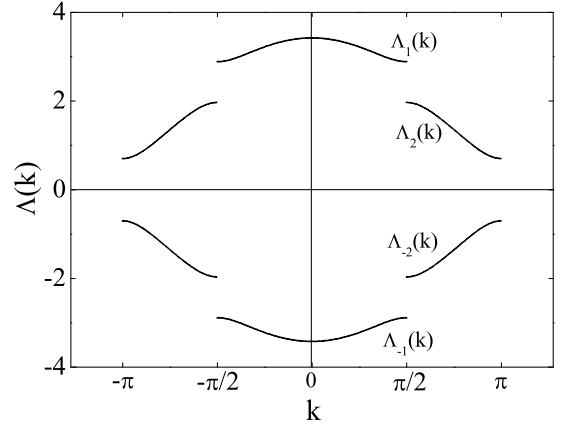} \caption{The representative fermionic excitation
spectra in the non-hole sector of the $S=1$ QIM with $Q>0$, $J_1 =
1.01$, $J_2 = 1$, $D_1 =1.82$, $D_2 =0.9$. This kind of excitations
for a sector with fixed hole number is similar though the
quasi-momentum is determined by the corresponding secular equation.}
 \label{fig:spectra12}
\end{figure}
It should be noticed that $\Lambda_{\pm r}(k)$ (for $r=1,2$) are
invariant under the shift $k\rightarrow k+\pi$. Thus the
quasi-momentum $k$ is constrained in one of the following regimes:
(1) $k \in[-\pi/2,\pi/2)$ or (2) $k\in[-\pi,-\pi/2)\cup[\pi/2,\pi)$.
Here, we choose $k$ to be in the first regime for $\Lambda_{\pm
1}(k)$ and in the second regime for $\Lambda_{\pm 2}(k)$.
Fig.~\ref{fig:spectra12} shows the typical energy dispersions for a
dimerized system with $Q>0$. The case with $Q<0$ can be obtained by
a reflection under the shift.

\subsection{The single-hole sector}\label{sub:n1}

In the $p=1$ sector, the hole breaks two bonds connecting to it. For
convenience, we assume that the hole is located at the site $L$. The
reduced Hamiltonian can be rewritten as
\begin{equation} \label{eq:h11}
H_{(1,L)}=H_{(0,L-1)}' - D_L ,
\end{equation}
where $H_{(0,L-1)}'$ is defined by Eq.~(\ref{eq:h03}) but with the
open boundary conditions. The procedure for diagonalizing
$H_{(0,L-1)}'$ is the same as for $H_{(0, L)}$. The eigen
wavefunction is determined by Eq.(\ref{eq:psi}) in a similar way.
The only difference is that now the $M$-matrix is a $(L-1)\times
(L-1)$ matrix, with $M_{L-1,1}=M_{1,L-1}=0$, and $M_{1,1}=a_0(\equiv
D_1^2)$.

Due to the open boundary condition, the Ansatz
Eq.~(\ref{wavedimerPBC}) is no-longer valid. Intuitively, a
reflection wave ($e^{-ikj}$) will be induced at the boundary in
addition to the incoming wave ($e^{ikj}$). Thus we suggest another
Ansatz for the dimerized open chain as follows:
\begin{eqnarray} \label{wavedimerOBC}
\begin{split}
(\Phi_k)_{2j}~~~&= A_{e}\left( e^{2i k j} - t_{e} e^{-2i k j} \right) ,\\
(\Phi_k)_{2j+1}&= A_{o}\left( e^{i k (2j+1)}- t_{o} e^{-i k(2j+1)}
\right) .
\end{split}
\end{eqnarray}
With this Ansatz, we find that $\tau$ and $\Lambda(k)$ take the same
form as those defined in the $p=0$ case. The reflection
coefficients, $t_o$ and $t_e$, are given by
\begin{eqnarray}
t_e=\frac{e^{iLk}}{e^{-iLk}},~~
t_o=\frac{(b_1e^{-ik}+b_2e^{ik})e^{iLk}}{(b_1e^{ik}+b_2e^{-ik})e^{-iLk}}.
\end{eqnarray}
Moreover, the quasi-momentum $k$ is now determined by the following
secular equation
\begin{eqnarray}
&&2b_2\left[b_1^2+b_2^2+2b_1b_2\cos 2k\right]\sin Lk\nonumber\\
&&=(a_0-a_1)\left[b_1\sin(L-2)k+b_2\sin Lk\right]\nonumber\\
&&~~~~\times\left[(a_1-a_2)\pm W\right]. \label{k-dimer}
\end{eqnarray}
The equation is symmetric under $k \rightarrow -k $, thus we only
need to solve the complex or positive $k$'s. The solution can be
further simplified if $b_1=b_2$ ( or $D_1J_1=D_2J_2$) where $t_o =
t_e$.

\subsection{The multi-hole sectors}\label{sub:nmany}

The previous approach is extended to the subsystems or sectors with
more holes. The holes break the Ising couplings, leading to a series
of segments of the spin-1/2 TIM's. For the periodic chain, the
number of these segments is equal to the number of holes $p$
(including the segment of zero length). If the holes are located at
$(x_1,x_2,\cdots,x_p)$, then the corresponding reduced Hamiltonian
is given by Eq.(\ref{eq:h0}), i.e., $H_{(p,L)}
  =\sum_{n=1}^{p}H^{'}(0,l_n)-\sum_{n=1}^{p}D_{x_n}$,
with $l_n=x_n-x_{n-1}-1$(here, $x_0=x_p-L$) being the length of the
$n$-th segment and
\begin{eqnarray}
H^{'}(0,l_n)& =&-\sum_{j=x_{n-1}+1}^{x_{n}-2} J_{j,j+1}
\sigma_{j}^z\sigma_{j+1}^z \nonumber \\
&&-\frac{1}{2}\sum_{j=x_{n-1}+1}^{x_{n}-1} D_j
\sigma_{j}^x-\sum_{j=x_{n-1}+1}^{x_{n}-1}\frac{D_j}{2}.
\end{eqnarray}

Each segment Hamiltonian $H'(0,l_n)$ can be diagonalized as
$H^\prime (0,l_n) = \sum_{k} \Lambda(k)\left( \eta_k^\dagger
\eta_k-\frac{1}{2} \right)-\sum_{j=x_{n-1}+1}^{x_{n}-1}\frac{D_j}{2}
$, where $k$ is the quasi-momentum satisfying the secular equation
Eq.~(\ref{k-dimer}) (by replacing $L$ by $l_n$). Notice that
associated with the fixed length of the segment there are four
different kinds of configurations, depending on whether the two edge
holes are located at odd or even sites.

\section{The low energy spectra}\label{sec:excitation}

\subsection{Fermionic excitations in the non-hole sector}

In the previous section, we show how to diagonalize the reduced
Hamiltonians of different sectors. In a given hole sector, there are
four branches of quasi-particle excitations, given by
Eq.~(\ref{Lamb1}). Obviously, the two negative branches
$\Lambda_{-r}(k)$~($r=1,2$) will be filled in the ground state of
that sector.

In the $p=0$ sector, the ground state is given by $|\psi_0
\rangle=\prod_{kk^\prime} \eta^{\dagger}_{-1,k}
\eta^{\dagger}_{-2,k^\prime}|0\rangle$, where $\eta^\dagger_{-r,k}$
($r=1,2$) are the fermionic quasi-particle operators in the branches
$\Lambda_{-r}(k)$, $k$ and $k^\prime$ are the allowed momenta for
$r=1,2$, respectively. The ground state energy is given by $
E_0(p=0,L)=\frac{1}{2}\sum_{k}\Lambda_{-1}(k)+
\frac{1}{2}\sum_{k^\prime}\Lambda_{-2}(k^\prime)$.

The low energy excitations can be obtained by applying operators
$\eta_{2,k^\prime}^\dagger$ or $\eta_{-2,k^\prime}$ on the ground
state. The excitation energy is given by
$\Lambda_2(k^\prime)$($=-\Lambda_{-2}(k^\prime)$). When $Q>0$, the
energy gap between the lowest excitation and the ground state is
given by
\begin{eqnarray} \label{eq:Delta0}
 \Delta_0 = \sqrt{\Gamma^2 -\sqrt{\Gamma^4 -\delta^4_{-}}},
\end{eqnarray}
where $ \delta_\pm =\sqrt{|D_1D_2\pm 4J_1J_2|} $. If
$|D_1|=|D_2|=|D|$ and $|J_1|=|J_2|=|J|$, one has
$\Delta_0=2|J||\frac{1}{|\lambda|}-1|$ with $\lambda = 2J/D$,
reproducing the result of the uniform chain. In this case, the
branches 1 and 2 connect smoothly. In the presence of dimerization,
however, the two branches will split and produce a dimerization gap
\begin{eqnarray}
\Delta_d=\frac{1}{2}\left(\sqrt{\Gamma^2 + \sqrt{\Gamma^4
-\delta_{+}^4}}- \sqrt{\Gamma^2 -\sqrt{\Gamma^4
-\delta_{+}^4}}\right).
\end{eqnarray}

\subsection{Single hole excitation}

The ground state of the $p=1$ sector has $L$-fold degeneracy in the
uniform chain, because the energy does not depend on the position of
the hole. But in the presence of dimerization the degeneracy will be
$L/2$-fold. While the momentum-dependence of the spectra can be
determined as in the $p=0$ sector, the quasi-momentum $k $ must
satisfy the secular equation associated with the open chain of
length $L-1$.

In the lowest state of the $p=1$ sector, the bands with negative
energies are fully filled. The corresponding energy is then given by
$ E_0(1,L)=\frac{1}{2}[\sum_{k} \Lambda_{-1}(k)+ \sum_{k^\prime}
\Lambda_{-2}(k^\prime)] -\frac{1}{2}\max{(D_1,D_2)}$. The last term
is contributed from the hole which may locate at either even or odd
sites.

\begin{figure}[t]
\includegraphics[width=0.90\columnwidth, bb=0 0 321 233]
{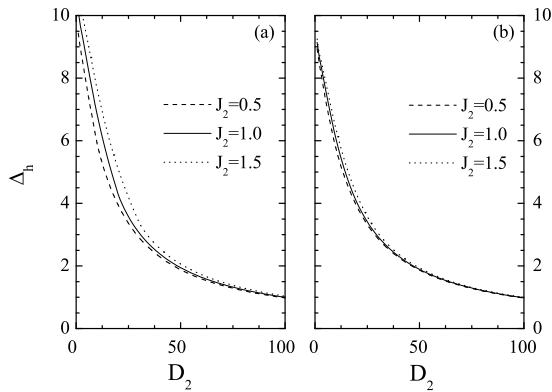} \caption{The hole excitation gap $\Delta_h$
as a function of $D_2$ for the system with periodic (left panel) or
open (right panel) boundary conditions. Three curves corresponding
to $J_2=0.5, 1.0$ and $1.5$ are plotted and $J_1=10$ and $D_1=2.0$
are fixed.} \label{dimerholegap}
\end{figure}

According to Theorem 2, the ground state of the original Hamiltonian
should be the one in the $p=0$ sector. So in addition to the
fermionic excitations, inclusion of the $p=1$ sector (adding a hole
to the system) will induce the hole excitations. The minimal hole
excitation gap is given by
$\Delta_h^{(1)}\equiv\Delta_h=E_0(1,L)-E_0(0,L)$. The typical
behavior of $\Delta_h$ (as a function of $D_2$) is shown in
Fig.~\ref{dimerholegap} for fixed $J_1=10$, $D_1=2.0$, and $L=2000$,
with either periodic or open boundary conditions.

In order to understand the difference in $\Delta_h$ for the periodic
and open chains, we consider a special case where the chain is
uniform and at the critical point, i.e., $D_1=D_2=D$, $J_1=J_2=J$,
and $\lambda=2J/D=1$. In this simple case, the energy spectrum is
given by $ \Lambda(k)=-|D|\sqrt{1+\lambda^2+2\lambda\cos{k}}$, and
the allowed quasi-particle momenta are $k=\frac{2m\pi}{L}$, $
(m=-L/2,\cdots,L/2-1)$ for the period boundary condition, and $k=
\frac{2m\pi}{2L+1}$, $(m=1,\cdots,L)$ for the open boundary
condition. Thus for sufficiently large $L$ we obtain
$\Delta_h\approx 0.136D$ and $\Delta_h\approx 0.318D$ for the open
and periodic chains, respectively. Their difference,
$\epsilon_s=0.182D$, does not change for larger $L$. So it is the
surface energy cost to turn into an open chain.

\subsection{Multi-hole Excitations}

It is straightforward to extend the above discussion to a multi-hole
system. Let us start with $p=2$. The eigenfunction in the $p=2$
sector is a direct product of the wavefunctions for the two $p=1$
segments, but with smaller lattice lengths $x-1$ and $L-x-1$,
respectively, here $x\equiv x_2-x_1$ is the distance between the two
holes sited at $x_1$, $x_2$.

For the uniform chain, the lowest energy state of the $p=2$ sector
should correspond to the configuration where two holes are close
together, as in this case the total surface energy is minimized. So
the lowest hole excitation gap in the $p=2$ sector is given by
$\Delta_h^{(2)} = E_0(2,L)-E_0(0,L)$. For $p>2$, one can further
show that the lowest state in the $p$-th sector is the configuration
in which all holes are close to each other. This is the hole
condensation phase in one dimension. The remain spin chain has a
length $L-p$. When $p<<L$ and $L\rightarrow \infty$, the finite size
effect is negligible so one has $\Delta_h^{(p)}\approx p\Delta_h$.

\subsection{Phase Diagram}\label{sec:QPT}

We now discuss how dimerization influence the phase diagram. We have
shown that the ground state still lies in the $p=0$ sector in the
presence of dimerization. However, dimerization splits the bands and
induce a gap at higher energies. The quantum critical points extend
to lines, which are determined by the gapless condition, i.e.
$\Delta_{0}=0$, or according to Eq.~(\ref{eq:Delta0}),
\begin{eqnarray}
D_1D_2=\pm 4J_1J_2.
\label{critical}
\end{eqnarray}

\begin{figure}[t]
\includegraphics[width=0.85\columnwidth, bb=10 10 284 221]{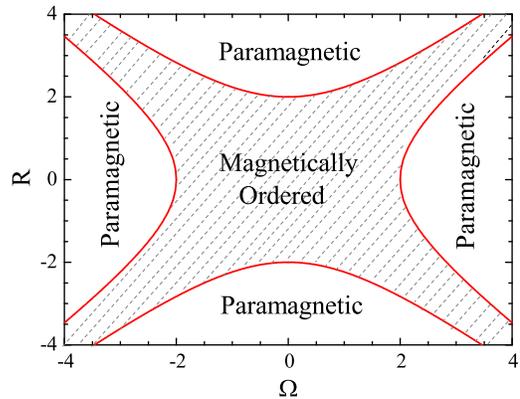}
\caption{(Color Online) Ground state phase diagram of the $S=1$ QIM.
The $\Omega$-axis and $R$-axis describe the dimerization strength
and the competition parameter, see Eq.(\ref{RO}) in the main text.
The dashed area is the magnetic ordered phase which extends slightly
for small $\Omega$ but split into two parts for $|\Omega|>2$.}
\label{fig:phase}
\end{figure}

The influence of dimerization can be seen more clearly by
introducing ( assuming $J_1J_2>0$)
\begin{eqnarray}\label{RO}
R=\frac{D_1+D_2}{2\sqrt{J_1J_2}},
~~\Omega=\frac{D_1-D_2}{2\sqrt{J_1J_2}}.
\end{eqnarray}
where $R$ describes the competition between $D_j$ and $J_j$, and
$\Omega$ is a measure of the dimerization strength. In the uniform
chain limit, $R=2/\lambda$, $\Omega=0$. With increasing $\Omega$, we
have the following three situations:

(1) In the weak dimerization regime, $|\Omega|<2$, there is a pair
of symmetric critical points, $\pm R_c$, with $
R_c=\sqrt{\Omega^2+4}$. The ground state is magnetically ordered
when $|R|<R_c$.

(2) In the strong dimerization regime, $|\Omega|>2$, there are two
pairs of symmetric critical points, $\pm R_{c_1}, \pm R_{c_2}$, with
$ R_{c_1}= \sqrt{\Omega^2+4}, R_{c_2}=\sqrt{\Omega^2-4}$. The
magnetically ordered phase appears when $R_{c_2}<R<R_{c_1}$.

(3) When $|\Omega|=2$~($R_{c_2}=0$), there are three critical
points, which take values $\pm 2\sqrt{2},0$ respectively. The ground
state is magnetically disordered when $|R|>2\sqrt{2}$, but ordered
(either ferromagnetic or anti-ferromagnetic, depending on the signs
of $J_{1,2}$) when $|R|<2\sqrt{2}$. Note that the $R=0$ point
corresponds to an alternating (or staggered) array of single-ion
anisotropy $(D,-D,...,D,-D)$. It becomes critical when
$D=\pm2\sqrt{J_1J_2}$.

The $R-\Omega$ ground state phase diagram is plotted in
Fig.~\ref{fig:phase}. The magnetic ordered phase and paramegnetic
disordered phase are separated by the critical lines (red).

\section{Thermodynamic properties}
\label{sec:thermal}

In this section, we study the thermodynamic properties of the $S=1$
QIM. As was shown previously, this model is exactly solvable not
only for the ground state, but also for all excited states. However,
exactly evaluating the thermodynamic quantities is still a very hard
task, particularly for large system size and dimerization. Here, we
shall develop the recursion method proposed in Ref.~\cite{Yang} in
the evaluation of the partition function as well as other
thermodynamic quantities for either uniform and dimerized chains.

\subsection{Recursion method}

Based on the HDS, the partition function $Z(L)$ of the system with
lattice length $L$ can be expressed as the sum of all the partition
functions of the subsystems, i.e.,
\begin{equation}
Z(L)=\sum_{p=0}^L Z(p,L).
\end{equation}
Where, $Z(p,L)$ is the partition function of the $p$-th sector.
Because for fixed $p$, there are many different hole configurations.
So $Z(p,L)$ can be further rewritten as a sum over all possible hole
configurations
\begin{equation}
Z(p,L)=\sum_{\{x_1\cdots x_p\}}Z ( x_1,\cdots, x_p).
\end{equation}

For each hole configuration $\{x_{1},\cdots, x_{p}\}$, the
corresponding partition function of the open chain is given by
($\beta=1/k_BT$)
\begin{eqnarray}
Z(x_1,\cdots,x_p)=\mathrm{Tr} e^{[
-\beta(\sum_{n=1}^{p+1}H^\prime(0,l_n)-\frac{1}{2}
\sum_{n=1}^pD_{x_n})]}.
\end{eqnarray}
Where, $H^\prime(l_n)$ is the Hamiltonian of the n-th segment.

The partition functions of each segments can be regarded as the
building blocks of total partition of the original system. These
building blocks are denoted by $z(l_n)$, the partition functions of
the spin-1/2 TIM segments with length $l_n$. Then, in the uniform
case, $Z(L)$ can be expressed as $
Z(L)=\sum_{p=0}^L\sum_{\{l_n\}}z(l_1)\alpha z(l_2)\alpha\cdots
\alpha z(l_{p+1})$,  where $\alpha=\exp(\beta D/2)$ is the partition
function of a hole, $z(1)=2\cosh(\beta D/2)$, and $z(0)\equiv 1$. As
the length of the allowed segment may vary, one has the summation
constraint $\sum_{n=1}^{p+1}l_n=L-p$. If we denote
$Z^{(p)}(L-p)\equiv z(l_1)z(l_2)\cdots z(l_{p+1})$, then the total
partition function can be rewritten as
\begin{eqnarray}
\label{partition} Z(L)=\sum_{p=0}^L\alpha^p Z^{(p)}(L-p).
\end{eqnarray}
Where, $\alpha^p$ is contributed from the hole's.

To numerically evaluate the partition function, it is practically
convenient to use the following recursion formula:
\begin{eqnarray}
\label{partition} Z^{(p)}(l)=\sum_{j=0}^l z(j) Z^{(p-1)}(l-j),
\end{eqnarray}
where, $Z^{(-1)}(l)\equiv \delta_{l,0}$ and $Z^{(0)}(l)\equiv z(l)$.
By this way, we firstly calculate the building block, $z(l_n)$, and
then by iterative use of above relation, evaluate the partition
function of the $S=1$ system. We find that this method is particular
efficient for the uniform chain, where the system size $L$ could be
as large as $L=10000$.

It is non-trivial to extend the above recursion method to the spin
chain in the presence of dimerization. Here, there are four kinds of
blocks, associated with the parity of the two ends. Thus we can
denote them by $z_{r_1r_2}(l_n)$, with $r_{1,2} (=o,e)$ indicating
the left/right ends respectively. The $S=1/2$ TIM segments with odd
or even end sites can be solved exactly. The analytical expressions
for $z_{r_1r_2}(l_n)$ will be provided in a separated supplementary
material\cite{EPAPS}. In the following, we present some numerical
results obtained by the recursion method while the system size is
kept at $L=2000$. For simplicity, we use the open boundary condition
for the original $S=1$ chain. The extension to the periodic boundary
condition is straightforward.

\subsection{The uniform spin chain}
Thermodynamics quantities can be calculated from the partition
function. In our model, an important physical quantity is the
thermal average of the hole number, defined by
\begin{eqnarray}
N_h=\frac{1}{Z(L)}\sum_{p=0}^L p\alpha^pZ^{(p)}(L-p). \label{nhole}
\end{eqnarray}

\begin{figure}[b]
\includegraphics[width=0.93\columnwidth, bb=13 13 342 242]
{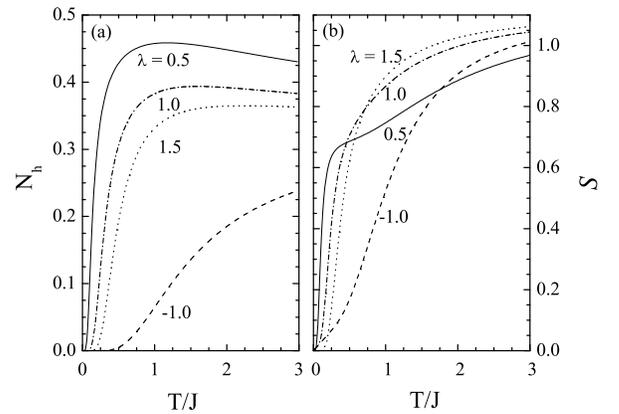} \caption{Temperature dependence of the hole
concentration $N_h$ (a) and the entropy ${\cal S}$ (b) for a uniform
spin chain with $\lambda=-1,0.5,1,1.5$, respectively.}
\label{fig:holeuniform}
\end{figure}

Fig.~\ref{fig:holeuniform}(a) shows the temperature dependence of
$N_h$ for several $\lambda$. At low temperatures, $N_h$ increases
rapidly with increasing temperature for small and positive
$\lambda$. The proliferation of the hole number at low temperatures
is obviously due to the smallness of the hole excitation gap.
However, for the negative $\lambda$, say, $\lambda=-1$, the hole
excitation gap is relatively larger, so $N_h$ increases much slowly
with temperatures.

Another interesting physical quantity is the entropy ${\cal S}$,
which we plot as a function of temperature for several different
$\lambda$ in Fig.~\ref{fig:holeuniform}(b). We find that the
suppression of the entropy ${\cal S}$ is stronger for larger
$\lambda$. The suppression is even more pronounced for the negative
$\lambda$. These behavior are similar to the temperature dependence
of the hole number and are also due to the hole excitation gap.

We also find that the $N_h$ approaches about $1/3$ in the high
temperature limit for all $\lambda$. Correspondingly, the entropy
saturates at the value $\ln 3$ in the high temperature limit(not
fully shown in Fig.~\ref{fig:holeuniform}).

\subsection{The dimerized spin chain}

\begin{figure}[ht]
\includegraphics[width=0.93\columnwidth, bb=13 13 294 220]
{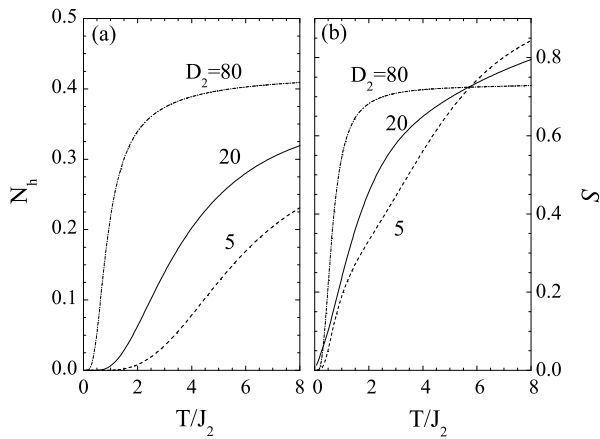} \caption{(a) Temperature dependence of the hole
concentration $N_h$ and (b) the entropy ${\cal S}$  for a dimerized
spin chain in three representative cases: $D_1D_2=J_1J_2$ (dashed
line), $4J_1J_2$ (real line), $16J_1J_2$ (dash-dotted line),
respectively. Where, $J_1=10$, $J_2=1$, and $D_1=2$ are fixed.}
\label{fig:holedimer}
\end{figure}

\begin{figure}[h]
\includegraphics[width=1.0\columnwidth, bb=14 14 321 240]
{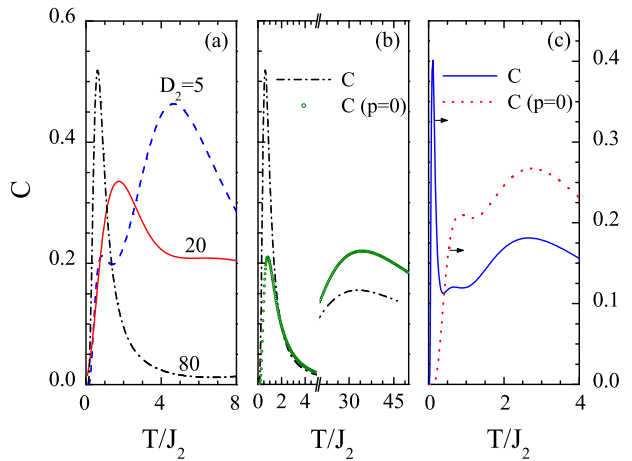} \caption{(Color Online) (a) The specific heat C for
$D_1D_2=J_1J_2$, $D_2=5$ (blue-dashed line), $D_1D_2=4J_1J_2$,
$D_2=20$ (red-real line), $D_1D_2=16J_1J_2$, $D_2=80$ (dash-dotted
line) with $J_1 =10$, $J_2=1$, and $D_1=2$, respectively; (b) The
specific heat C (dash-dotted line) and the corresponding specific
heat C(p=0) for the $p=0$ subsystem (green open circle) for
$D_1D_2=16J_1J_2$ and $J_1=10$, $J_2=1$, $D_1=2$, $D_2=80$,
respectively. (c) The specific heat $C$ (blue-real line) and C(p=0)
(red-doted line) with $J_1=J_2=1$, $D_1=2$, and $D_2=7$.}
\label{fig:heatdimer}
\end{figure}

The recursion method, after some extensions discussed previously, is
also used to evaluate thermodynamic quantities for a dimerized
system. Figs.~\ref{fig:holedimer} and \ref{fig:heatdimer}(a) show
the temperature dependence of the entropy ${\cal S}$, the average
hole concentration $N_h$, and the specific heat $C$ in the ordered
($D_1D_2=J_1J_2$), critical ($D_1D_2=4J_1J_2$), and disordered
phases($D_1D_2=16J_1J_2$), respectively. Here we fix $J_1=10$,
$J_2=1$, $D_1=2$, and choose several $D_2$ (= $5$, $20$, and $80$)
in Figs.~\ref{fig:holedimer} and \ref{fig:heatdimer}(a). Note that
both the dimerization strength and competition parameter are tuned
by varying $D_2$. The general features of the hole number and the
entropy are similar to that in the uniform case as the hole
excitation gap plays the role at low temperatures (note that when
$J_1=10$, $J_2=1$, and $D_1=2$, $\Delta_h=7.450$, $3.902$, and
$1.202$ for $D_2=5$, $20$, and $80$, respectively).

In Figs.~\ref{fig:heatdimer}, the specific heat is plotted as a
function of temperature in several cases. We find that at
low-temperatures the specific heat is peaked at the temperature
scale near the hole excitation gap. The peak behavior changes
depending on the competition of the fermionic and hole excitations.
The influence of the dimerization on the specific heat can be
clearly seen at relatively higher temperatures, as the dimerization
induced gap is much larger than the hole or fermion excitation gaps.
Fig.~\ref{fig:heatdimer}(b) shows the specific heats of the $S=1$
QIM system and its $p=0$ sector for the case
$J_1=10,~J_2=1,~D_1=2,~D_2=80$. Note that the $p=0$ sector is
identical to the corresponding $S=1/2$ TIM. In this case
$\Delta_0(\sim\Delta_h)$ is very small, leading to a sharp peak in
the $S=1$ QIM system. We also find a well separated and relatively
round peak in the higher temperature regime where the energy scale
is close to the dimerization gap $\Delta_d(=40.025)$.

The competitions among the dimerization effect, the hole excitations
and the fermion excitations can be seen more clearly from
Fig.~\ref{fig:heatdimer}(c) where $J_1=J_2=1$, $D_1=2$, and $D_2=7$.
In this case, the three kinds of gaps are well separated,
$\Delta_h<\Delta_0<\Delta_d$, so that the specific heat exhibits
three peaks. The first one is a sharp peak around $T\sim \Delta_h$
\cite{explaingap}, other two peaks are around $T\sim \Delta_0$ and
$T\sim \Delta_d$, respectively. By contrast, the low temperature
sharp peak disappears in the $p=0$ sector and other two peaks still
persist [see the red dotted line in Fig.~\ref{fig:heatdimer}(c)].
This is in agreement with the fact that no hole excitation exists in
the $p=0$ sector.

\begin{figure}[t]
\includegraphics[width=0.93\columnwidth, bb=13 13 300 230]
{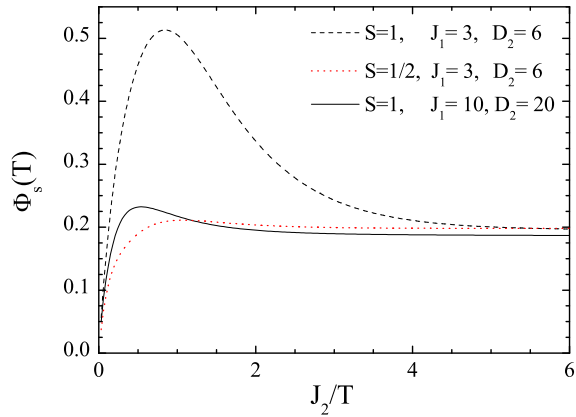} \caption{(Color Online) The scaled free energy
for the dimerized systems in the quantum critical phase where
$D_1D_2=4J_1J_2$. Two typical situations are plotted: ($J_2=3$,
$D_2=6$) and ($J_2=10$, $D_2=20$), with small and large hole
excitation gaps, respectively. Other parameters are $J_2=1$ and
$D_1=2$. As a comparison, the corresponding $S=1/2$ TIM case is also
plotted.} \label{fig:phidimer}
\end{figure}

We also calculated the scaled free energy $\Phi_s(T)$ at the
critical point $D_1D_2=4J_1J_2$ for the dimerized spin chain.
$\Phi_s(T)$ is defined by
\begin{equation}\label{eq:phis}
\Phi_s (T) =\frac{2|J_2| [F(0) -F(T)]}{T^2}.
\end{equation}
This quantity, which is identical to the specific heat coefficient
at low temperatures, was introduced in Ref.\cite{Kopp} in order to
show the temperature persistence of the quantum critical scaling
behavior in the $S=1/2$ TIM. In the critical region where quantum
critical fluctuations dominate this quantity should be a constant.
It was shown that in the $S=1/2$ uniform TIM it deviates from the
constant only when $T\gtrsim J/2$, indicating a rather higher
temperature scale below which the quantum critical scaling behavior
persists \cite{Kopp}. In our recent work Ref.\cite{Yang}, we found
that in the uniform $S=1$ QIM this behavior is strongly suppressed
by the hole excitations. Here, we find that the similar conclusion
can be inferred in the presence of dimerization. In
Fig.~\ref{fig:phidimer}, we plotted the results for two cases (i)
$J_1=3$, $J_2=1$, $D_1=2$, $D_2=6$, and (ii) $J_1=10$, $J_2=1$,
$D_1=2$, and $D_2=20$. As a comparison, the corresponding result for
the dimerized $S=1/2$ TIM is also plotted. We find that the quantum
critical scaling behavior at $T\rightarrow 0$ (or $J_2/T\rightarrow
\infty$ in the Figure \ref{fig:phidimer}) persists approximately at
finite $T\approx 0.5 J_2$ for the dimerized $S=1/2$ TIM and the case
(ii), and at $T\approx 0.2 J_2$ for the case (i), respectively. This
is because that the hole gap in the case (i) is much smaller than
the case (ii). Consequently, the hole excitations play more
significant role in suppressing the quantum critical scaling
behavior in the former case. This result is consistent with the
conclusion in Ref. \cite{Yang}.

\subsection{Hole condensations: the case with finite $D_z$}

Now we turn to the case with non-zero $D_z$. It is straightforward
to show that the $D_z$-term plays the role of chemical potential for
holes in the $S=1$ QIM~\cite{Yang}. More specifically, the energy of
a $p$-hole sector with finite $D_z$ is related to that with $D_z=0$
by the following relationship [$E_0^{(p)}(0)\equiv E_0(p,L)$]:
\begin{equation}
E_0^{(p)}(D_z)=E_0^{(p)}(0)+ pD_z.
\end{equation}
Thus, with finite $D_z$, the fermion excitation spectra remain
unchanged, but the hole excitation gap becomes\cite{note2}
\begin{equation}
\Delta_h(D_z)=\Delta_h(0)+D_z. \label{eq:Dzgap}
\end{equation}

\begin{figure}[h]
\includegraphics[width=0.9\columnwidth, bb=14 14 297 227]
{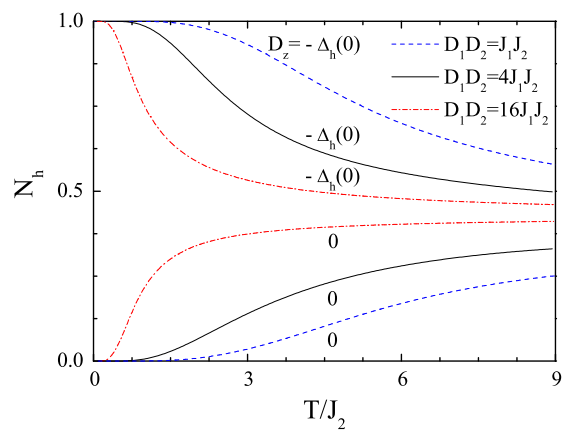} \caption{(Color Online) Temperature dependence
of the hole concentration $N_h$ for $D_z=0$ or $D_z = -\Delta_h(0)$.
Other parameters are $J_1=10$, $J_2=1$, and $D_1=2$.}
\label{fig:dimerhole2}
\end{figure}

Therefore, the ground state depends strongly on the value of $D_z$.
When $D_z>-\Delta_h(0)$, the hole excitation is positive, and the
ground state is still in the $p=0$-sector. But when
$D_z<-\Delta_h(0)$, the hole excitation gap is negative. This
indicates that the $p=0$ sector is no longer the lowest energy state
and there are holes in the ground state. As a result, Theorem 2 is
no longer valid in the present case. When $L$ is sufficiently large
and $p$ is relatively small, the lowest energy of the $p$-sector can
be approximated by $ E_0^{(p)}(0)\approx E_0^{(0)}(0)+p\Delta_h(0)$
as discussed in the previous section. Then, $ E_0^{(p)}(D_z)\approx
E_0^{(0)}(D_z)+p[\Delta_h(0)+D_z]$, so that we have
\begin{equation}
E_0^{(L)}(D_z)\lesssim \cdots\lesssim E_0^{(1)}(D_z)\lesssim
E_0^{(0)}(D_z).
\end{equation}
Thus, the order of the band structure hierarchy is completely
overturned. In this case, the ground state is in the $p=L$ sector
and all sites are occupied by holes\cite{Yang}. This can be also
seen clearly from Fig.~\ref{fig:dimerhole2}, where the temperature
dependence of the hole concentration $N_h$ is shown. We find that in
the zero temperature limit, $N_h$ is equal to 1 when $D_z\le
-\Delta_h(0)$ or 0 when $D_z > -\Delta_h(0)$.

\section{Summary}\label{sec:summary}

In this paper, we have studied a class of exactly solvable $S=1$
QIMs with single-ion anisotropy. They exhibit a hierarchy of the
band structure with both fermionic and hole excitations. The
hole excitation gap can be tuned by the longitudinal crystal field
$D_z$. It becomes zero when $D_{z}$ is equal to $-\Delta_h(0)$. The
ground state exhibits three distinct phases: the magnetically
ordered or disordered phases when $D_z> -\Delta_h(0)$, or the hole
condensation phase when $D_z< -\Delta_h(0)$. The transition to the
hole condensation phase is of the first order.

We have shown that dimerization does not destroy the exact
solvability of this model. To our knowledge, this is the first
example in dimerized $S=1$ quantum spin systems where all the eigen
states as well as the wavefunctions and the eigen energies can be
solved exactly. The hole excitations enhance the thermodynamic
fluctuations as evidenced in the specific heat which shows a sharp
peak in the low-temperature region where the hole excitations
proliferate. This strongly reduces the characteristic temperature
below which the quantum criticality persists. All these results are
robust against the dimerization. However, dimerization deforms the
phase diagram and affects the high energy behavior.

We have developed a recursion method to sum over all hole
configurations efficiently. This provides a powerful approach for
evaluating rigourously all thermodynamic quantities as well as
static and dynamic correlation functions of the QIMs in the
thermodynamic limit. The detailed derivations for these quantities
in the presence of dimerization will be provided as a supplementary
material\cite{EPAPS}. The recursion method holds not just for the
model studied here. It can be easily extended and applied in other
physical systems whose Hamiltonian can be written as a sum of
independent spin segments, separated by nonmagnetic impurities, such
as Pd- or Zn-doped quasi-one-dimensional antiferromagnets
Sr$_2$(Cu$_{1-x}$Pd$_x$)O$_3$ or
Cu$_{1-x}$Zn$_x$GeO$_3$\cite{Sirker,Hase}.

\section*{Acknowledgments}
Z.H.Y. would like to thank Z.X. Xu for helpful discussions.
This work was supported in part by the National Natural Science
Foundation of China, the national program for basic research of
China, the PCSIRT (IRT-0754) and SRFDP (No.J20050335118) of
Education Ministry of China.


\begin{thebibliography}{99}
\bibitem{note3}While "dimerization" usually refers to the alternating antiferromagnetic
coupling, here we use it to indicate that both the Ising coupling
and the single-ion anisotropy can be of alternating pattern with
period 2.
\bibitem{explaingap} In the presence of dimerization the hole excitation gap
is also dependent on the odd/even sites where the hole located. Here
we define the smaller one by $\Delta_h$.
\bibitem{note2}In the following, we use $\Delta_h(0)$ to explicity
denote the hole excitation gap in the case with $D_z=0$, in order to
distinguish the one with $D_z\neq 0$.
\end{thebibliography}
\end{document}